\documentclass[aps,twocolumn,prl,preprintnumbers,amsmath,amssymb,superscriptaddress,floats]{revtex4-2}

\usepackage{graphicx}
\usepackage{bm}
\usepackage{hyperref}

\usepackage{natbib}
\usepackage{braket}
\usepackage{float}
\usepackage{color}
\usepackage{nicefrac}

\begin{document}
\title{ THz Electrodynamics of Mixed-Valent YbAl$_3$ and LuAl$_3$ Thin Films}

\author{D. Barbalas}
\affiliation{Department of Physics and Astronomy, The Johns Hopkins University, Baltimore, MD 21218, USA}

\author{S. Chatterjee}
\affiliation{Laboratory of Atomic and Solid State Physics, Department of Physics,
	Cornell University, Ithaca, New York 14853, USA
}
\affiliation{Department of Electrical \& Computer Engineering,
	University of California, Santa Barbara, CA 93106, USA}
 
\author{D. G. Schlom}
\affiliation{Department of Materials Science and Engineering, Cornell University, Ithaca, New York 14853, USA}
\affiliation{Kavli Institute at Cornell for Nanoscale Science, Ithaca, New York 14853, USA}
\affiliation{Leibniz-Institut für Kristallzüchtung, Max-Born-Str. 2, 12489 Berlin, Germany}

\author{K. M. Shen}
\affiliation{Laboratory of Atomic and Solid State Physics, Department of Physics,
	Cornell University, Ithaca, New York 14853, USA
}
\affiliation{Kavli Institute at Cornell for Nanoscale Science, Ithaca, New York 14853, USA}

\author{N. P. Armitage}
\affiliation{Department of Physics and Astronomy, The Johns Hopkins University, Baltimore, MD 21218, USA}

\date{\today}

\begin{abstract}
We present THz measurements of thin films of mixed-valent YbAl$_3$ and its structural analogue LuAl$_3$. Combined with Fourier transform infrared (FTIR) spectroscopy, the extended Drude formalism is utilized to study the low-frequency transport of these materials. We find that LuAl$_3$ demonstrates conventional Drude transport whereas at low temperatures YbAl$_3$ demonstrates a renormalized Drude peak and a mid-infrared (MIR) peak in the conductivity, indicative of the formation of a mass-enhanced Fermi liquid (FL).  In YbAl$_3$ the extended Drude analysis demonstrates consistency with FL behavior below the FL coherence temperature $T^* < 40$ K with the scattering rate following $T^2$ proportionality and a moderate mass enhancement. Despite not observing a clear $\omega^2$ Fermi liquid-like frequency dependence the evidence is consistent with a moderate mass Fermi liquid, albeit one with a smaller mass than observed in single crystals. Furthermore, the extended Drude analysis also suggests a slow crossover between the FL state and the normal state above the $T^*$ in YbAl$_3$, indicative of incoherent hybridization effects persisting to high temperatures .

\end{abstract}

\maketitle

\section{Introduction}

Rare-earth intermetallic systems demonstrate a wide variety of low-temperature behavior due to the interplay between electronic correlations and local moment physics. At low temperatures, Kondo physics dominates and the screening of the $f-$electrons leads to the formation of a Fermi-liquid (FL) of heavy quasi-particles below a temperature $T^*$, giving their name to the heavy fermion (HF) class of materials. These HF materials demonstrate a wide variety of exotic physics including non-conventional superconductivity and quantum phase transitions \cite{Stewart1984,Coleman1987}. Related to the class of HF compounds are the family of mixed-valence (MV) materials which demonstrate weaker localization of the $f-$orbitals, allowing the interplay of both charge and spin degrees of freedom \cite{Varma1976,Fisk1992}. These materials are characterized by a significantly larger Kondo scale with respect to the Fermi liquid scale, i.e. $T_K >> T^*$. While the mixed-valent state is expected to smoothly evolve out of the HF state, there remains many unique phenomena that can arise. Due to their well-separated energy scales, these systems can be tuned to study a variety of emergent phenomena in strongly correlated materials. 

Generically, $f-$orbital systems are expected to exhibit qualitative behaviors consistent with the Anderson single-impurity model formulation, the generalization of the single impurity to the periodic Anderson model (PAM) has remained difficult to treat \cite{Bauer2004,Kumar2011}. The single-impurity model has had success in describing the high temperature state of many materials, but as additional coherence effects appear at the lowest temperatures the periodic Anderson model must be utitized to more accurately describe the apparent phenomena \cite{Brandow1986,Degiorgi1999,Georges1996,Coleman1987}. Based on the nature of the screening processes occurring, the PAM can give better describe how the crossover between the low-temperature Fermi liquid behavior and the normal state should occur. The concept of a slow crossover arises due to extended screening processes that exist due to the lattice of f-orbital ions that are not explicitly described in the single impurity model \cite{Tahvildar-Zadeh1997}. In particular, band structure effects and the relative magnitudes of the two energy scales $T^*$ and $T_K$ play a role in determining the temperature dependence of the system and help explain the variation observed in experimental results \cite{Burdin2009}. 

Despite extensive studies of lanthanide intermetallic compounds, uncertainty still exists as to why Yb-based correlated materials do not the show the same broad range of correlated phenomena as compared to other lanthanide series elements; most notably, there has historically been a stark absence of Yb-based materials that demonstrate unconventional superconductivity. Na\"{i}vely, the commonality between Ce ($4f^1$) and Yb ($4f^{13}$ in Yb$^{+3}$) compounds as $4f$ electron-hole counterparts would suggest that they would share a similar range of behaviors \cite{Fisk1992}; however, experimentally this has not been found to be the case \cite{Andres1975,Awasthi1993,Webb1986}. No Yb-based intermetallic material had demonstrated unconventional superconductivity until the discovery in $\beta-$YbAlB$_4$ with $T_c = 80$ mK by \citet{Nakatsuji2008} and it has remained the sole example. In $\beta-$YbAlB$_4$ the evidence of a non-FL state transitioning into a superconductivity state at very low temperatures is evidence that the system inherently resides near a quantum critical point driven by intrinsic valence fluctuations rather than external pressure \cite{Ramires2012}. This discovery of a historically elusive phenomena in Yb-based materials highlights the need for further study of the other mixed-valent Yb-compounds to better understand the role valence fluctuations in other strongly correlated systems.  

In YbAl$_3$, the Yb ion has been found to be mixed-valent with a temperature dependence average valence between Yb$^{2+}$ and Yb$^{3+}$ \cite{Havinga1973,Tjeng1993} and a large separation of energy scales between the heavy Fermi liquid (FL) coherence temperature $T^*$ = 34 - 40K and the Kondo temperature $T_K \approx 670$ K \cite{Rowe2002,Cornelius2002}. Consistent with heavy FL quasiparticles forming at low temperatures, there exists a renormalized Drude peak due to the formation of heavy-quasiparticles and a mid-infrared peak corresponding to excitations between the hybridized conduction and $f-$electron bands \cite{Degiorgi2001,Okamura2004}. The existence of a moderately heavy FL state is supported by experimental results for the specific heat coefficient $\gamma = 40 $ mJ/mol K and the measured dHvA effective mass below 1.5K of $m^* = 15-30 m_e$ \cite{Cornelius2002,Ebihara2000,Ebihara2003,Okamura2004}. More recently, the growth and subsequent characterization of thin films of YbAl$_3$ and LuAL$_3$ using DC transport and ARPES has also uncovered that valence fluctuations drive a Lifshitz transition at 21 K by shifting the chemical potential \cite{Chatterjee2016,Chatterjee2017}. These recent results suggest that interplay of mixed-valence and Kondo physics can lead to previously unexpected low-temperature behaviors which deserve closer study. In this work we present the optical response of thin films of YbAl$_3$ and their structural analogue LuAl$_3$ characterized by time-domain THz spectroscopy (TDTS) and Fourier-transform IR spectroscopy (FTIR). The low-energy scale accessible by our THz spectral range of 0.1-8 meV allows for the direct characterization of low-energy excitations of  hybridized states as a function of both frequency and temperature~\cite{Bosse2012,Bosse2016}. Using the extended Drude formulism, we observe that the quasiparticle scattering rate shows temperature scaling consistent with Fermi liquid theory in YbAl$_3$ whilst LuAl$_3$ demonstrates behavior consistent with a simple non-interacting metal. We also observe the formation of a moderately heavy FL in YbAl$_3$ with a strong temperature-dependent relative mass enhancement from 5 K to room temperature. Additionally, both the quasiparticle scattering rate and effective mass also demonstrate a slow temperature-dependent crossover suggestive of incoherent hybridization effects of Yb $f-$electrons to the Al conduction band above $T^* = 37$ K and persisting up to 100 K. 

\section{Experimental Details}
\begin{figure}
\begin{center}
	\includegraphics[width= 1.0 \columnwidth]{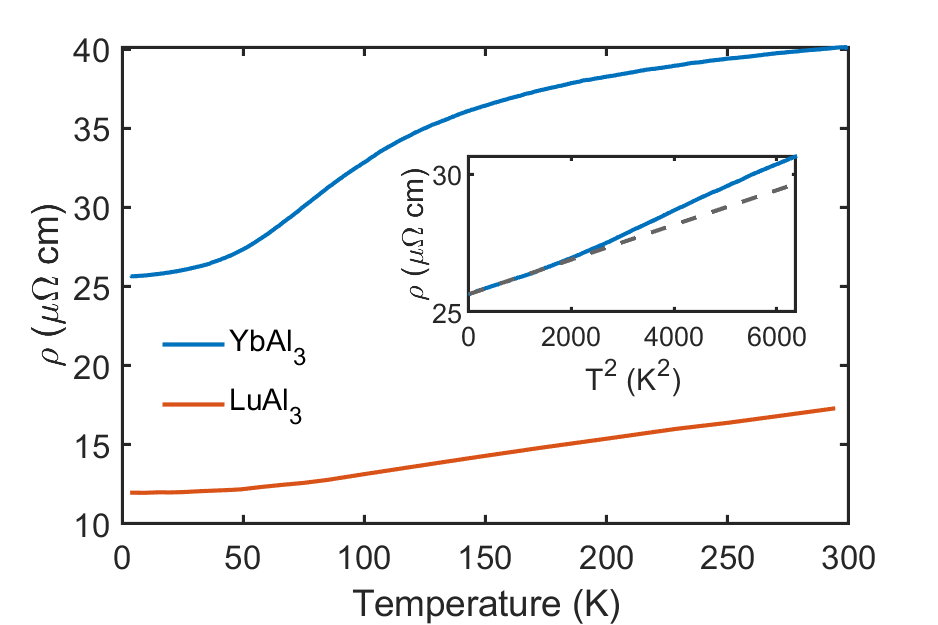}
	\caption{DC resistivity as a function of temperature for the YbAl$_3$ and LuAl$_3$ thin films.  These dc measurements were performed on similarly prepared, but different films than the THz measurements, so the correspondence with the THz data in Fig. 2 is not precise. }
	\label{resistivity}
\end{center} 
\end{figure}

\begin{figure}
\begin{center}
	\includegraphics[width= 1 \columnwidth]{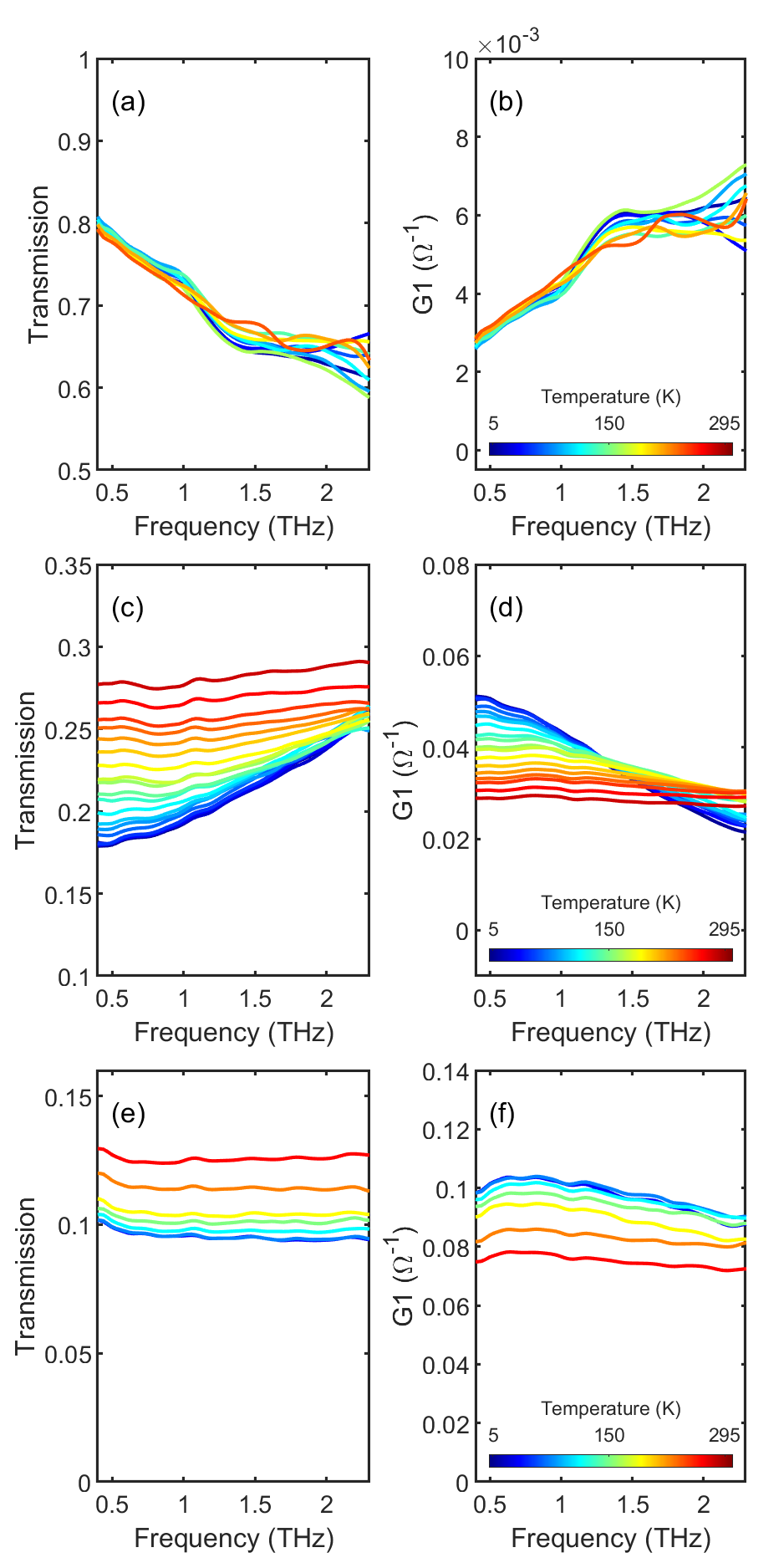}
	\caption{The raw THz transmission and the real conductance for (a),(b) the bare buffer layer, (c),(d) the YbAl$_3$ sample film on the buffer layer and (e),(f) the LuAl$_3$ sample film on the buffer layer. Note that the buffer layer has a small and temperature independent conductance. }
	\label{resistivity}
\end{center} 
\end{figure}

\begin{figure*}
	\begin{center}
		\includegraphics[width= 1.0 \textwidth]{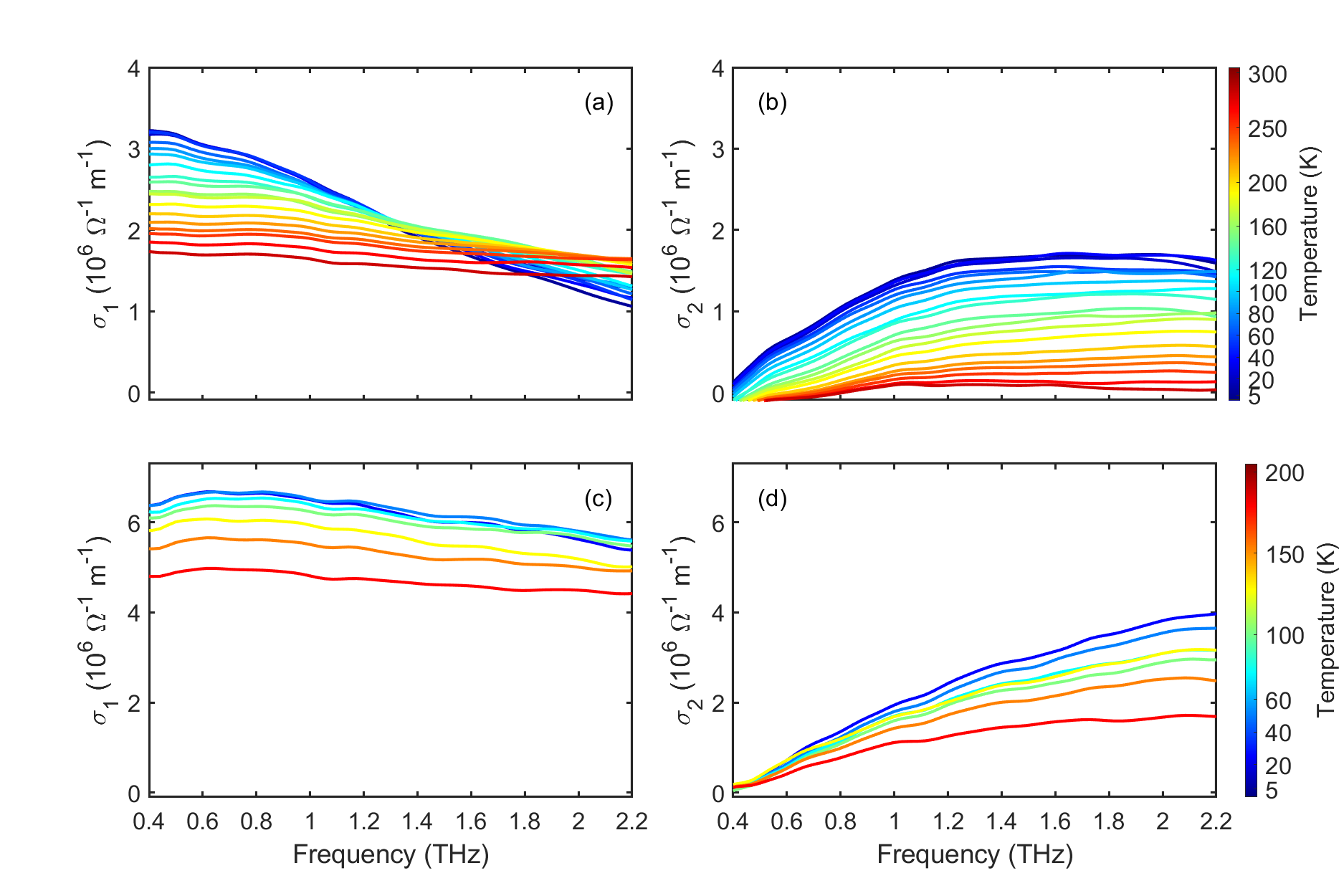}
		\caption{ (a), (b) Real and imaginary THz conductivity of YbAl$_3$ as a function of frequency, (c), (d)  Real and imaginary THz conductivity of LuAl$_3$ as a function of frequency. Note that the magnitude of the buffer layer conductance is less than 10\% of the sample films. }
		\label{THz_cond}
	\end{center}
\end{figure*}

\begin{figure*}
	\begin{center}
		\includegraphics[width= 1.0\textwidth]{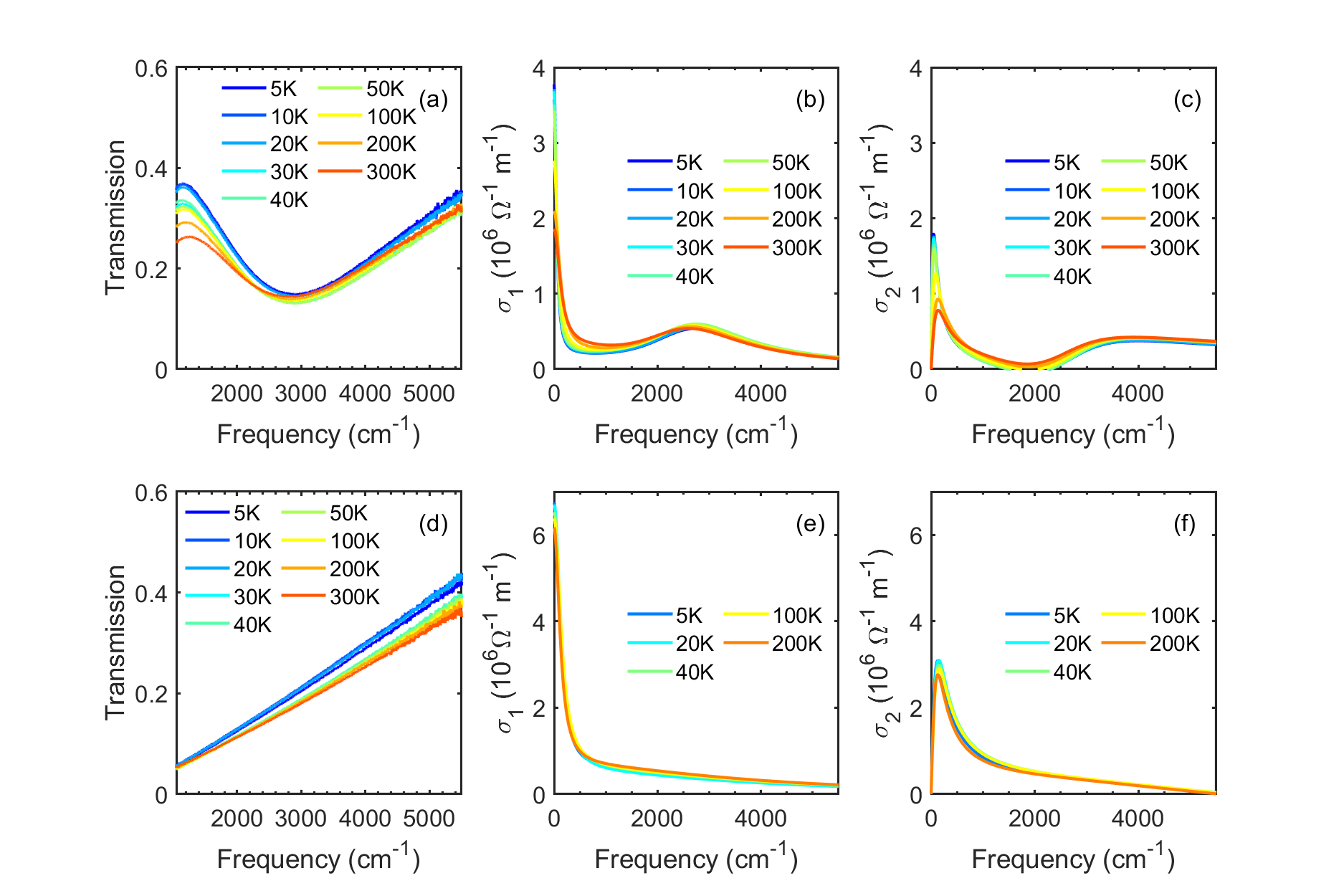}
		\caption{MIR transmission shown above for YbAl$_3$ (a) and LuAl$_3$ (d). The real and imaginary parts of the optical conductivity modeled by the combined THz complex conductivity and MIR transmission is shown below in (b),(c) for YbAl$_3$.   The real and imaginary parts of the optical conductivity for LuAl$_3$ are shown in (e) and (f).}
		\label{MIR}
	\end{center}
\end{figure*}
The (001)-oriented thin films of YbAl$_3$ and LuAl$_3$ used in this study were 15 nm thick and grown on (001) MgO substrates using molecular-beam epitaxy (MBE).  For all films, a 0.8 nm thick aluminum buffer layer was first deposited.  This allowed the growth of continuous, smooth films of LuAl$_3$/YbAl$_3$. In these studies, we investigated a 20 nm thick LuAl$_3$ film and a 15 nm thick YbAl$_3$ film.  The YbAl$_3$ was synthesized on top of a 5 nm thick LuAl$_3$ buffer layer on top of the Al buffer, which improved the quality of the YbAl$_3$ layer. All films were capped with 2 nm thick layer of Al which was assumed to oxidize and become insulating. Characterization of these films was conducted using x-ray diffraction, low energy electron diffraction and scanning transmission electron spectroscopy. Despite the dramatic advances in growing such films by MBE, it is important to note that the residual resistivity ratio (RRR = $\rho(\textrm{300 K})/\rho(\textrm{5 K})$\,) of the YbAl$_3$ film is still much less than the approximately 60 found in high quality single crystals~\cite{Ebihara2000}. More details about the thin film growth and characterization can be found elsewhere \cite{Chatterjee2016}.

The low-energy response of the films were characterized using time-domain THz spectroscopy (TDTS).  In TDTS, an approximately single-cycle ps long pulse of radiation is transmitted through the sample. One takes the Fourier transform of the transmitted electric field and references it to the signal through a bare MgO substrate to get the complex transmission of the thin films.  An advantage of the technique is that the complex conductivity can be directly extracted from the complex transmission without a Kramers-Kronig transform.  The complex conductivity $\widetilde{\sigma}$ can be determined from the transmission by,
\begin{equation}
	\widetilde{T}(\omega)= \frac{n_s + 1}{n_s + 1 + Z_0 d \: \widetilde{\sigma}(\omega)} e^{i  \frac{\omega}{c}\Delta L (n_s - 1)}   ,
\end{equation}
where $n_s$ is the substrate index of refraction, $Z_0$ is the impedance of free space, $d$ is the sample thickness, $\widetilde{\sigma}$ is the complex conductivity, $\omega$ is the frequency, $c$ is the speed of light, and $\Delta L$ is a correction factor due to the difference in thickness between the substrate and the sample \cite{Krewer2018}. In order to isolate the conductance of the LuAl$_3$/YbAl$_3$ samples from the buffer and capping layers, a reference film comprised of the buffer and capping layers (without a sample film) was used to correct for the presence of parallel conductance channels. The magnitude of the conductance of the reference film was less than 10\% of the conductance of the sample films and displayed no temperature dependence. By subtracting the conductance of the buffer layers from the total conductance of the sample films, the conductivity of the LuAl$_3$/YbAl$_3$ layer could be obtained. 

In tandem with the THz measurement, the mid-IR (MIR) response (400-7500 cm$^{-1}$, 0.05 - 930 meV) was studied using conventional Fourier-transform infrared (FTIR) spectroscopy down to 5 K. Transmission experiments were measured using a commercial FTIR spectrometer (Bruker Vertex 80V) using a Globar source and a solid-state DTGS detector. 

\section{Results} 

\begin{figure}
\begin{center}
    \includegraphics[width= \columnwidth]{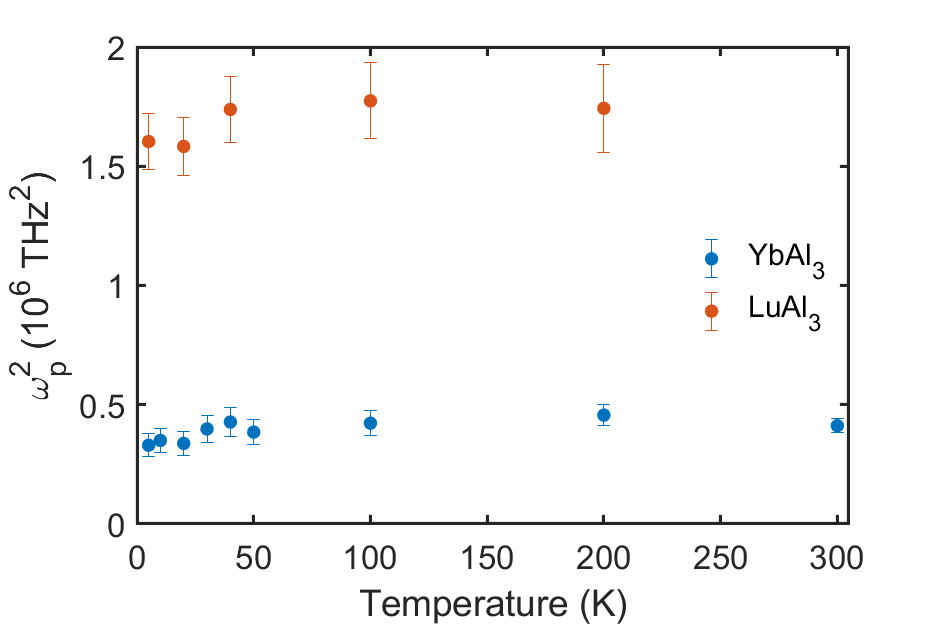}
	\caption{The intraband spectral weight as a function of temperature deduced from simultaneous fitting of the THz and MIR data for LuAl$_3$ and YbAl$_3$. The error bars are due to uncertainty in how spectral weight is divided amongst oscillators in the parameterization of the data. }
	\label{MIR_wp2}
	\end{center}
\end{figure}

In Fig.~\ref{resistivity}, the DC resistivity of films of the two materials are shown. At the lowest temperatures, the YbAl$_3$ film demonstrates a $\rho \propto T^2$ dependence as shown previously \cite{Chatterjee2016}. The resistivity saturates above some temperature $T^*$ that is of order 200 K.  The resistivity of the LuAl$_3$ film shows a monotonic decrease with decreasing temperature.

In Fig.~\ref{THz_cond} we present the real and imaginary parts of the THz conductivity for YbAl$_3$ and LuAl$_3$ at different temperatures. At high temperatures, both films demonstrate large scattering rates indicated by the flat, nearly frequency independent real conductivity.  Upon cooling the scattering rate decreases in both films; however, in YbAl$_3$, the decrease in the scattering rate is observed to be significantly greater. In the case of YbAl$_3$ sample, at the lowest temperatures there is an increase in the low-frequency spectral weight as seen by the enhancement of the real and imaginary parts of the conductivity compared to the room temperature state. This is consistent with the formation of a narrower Drude peak at the lowest temperatures. This is in contrast to the LuAl$_3$ film, which is characterized by a broader Drude-like contribution to the conductivity without a significant temperature dependence of the scattering rate. Additionally, from the real part of the conductivity it is clear that in LuAl$_3$ there is no significant change in the spectral weight, indicating the absence of any strong renormalization effects. 

The MIR transmission of the YbAl$_3$ and LuAl$_3$ thin films are shown in Fig.~\ref{MIR}. Using both the THz complex conductivity and the MIR transmission, the optical conductivity was modeled using Drude-Lorentz parameterization in the fitting software refFIT \cite{Kuzmenko2005}. The LuAl$_3$ data was fit by two Drude terms, whereas the fit to the YbAl$_3$ data used two Drude terms and a single high frequency Drude-Lorentz oscillator. In order to account for the buffer layer present in both sample films, a multilayer model was generated in refFIT using the parameterization of the bare buffer layer on the substrate to isolate the transmission of the individual sample layers. Due to the strong absorption of the underlying MgO substrate, the transmission below 1000 cm$^{-1}$ was not accessible. Hence, in the FIR region without experimental data, the modeled conductivity determined through the Kramers-Kronig consistent fitting can only give a rough guide, but is not a definitive measure of the conductivity.

From the MIR data on the YbAl$_3$ thin film, we see a clear MIR conductivity peak suggestive of a hybridization gap \cite{Marabelli1992},which becomes enhanced as the temperature decreases. The position of the MIR conductivity peak in YbAl$_3$ is centered around $\omega = 2800$ cm$^{-1}$; the corresponding value in single crystals has been found to be centered at $\omega \sim 2000$ cm$^{-1}$ \cite{Okamura2004, Lv2018}. While there is no strain expected to arise from the underlying Al and LuAl$_3$ buffer layers on MgO, this shift in the conduction electron-$4f$ hybridization resonance can be attributed to the effects of strong impurity scattering \cite{Coleman1987a}. 
\begin{figure}[h]
		\includegraphics[width= \columnwidth]{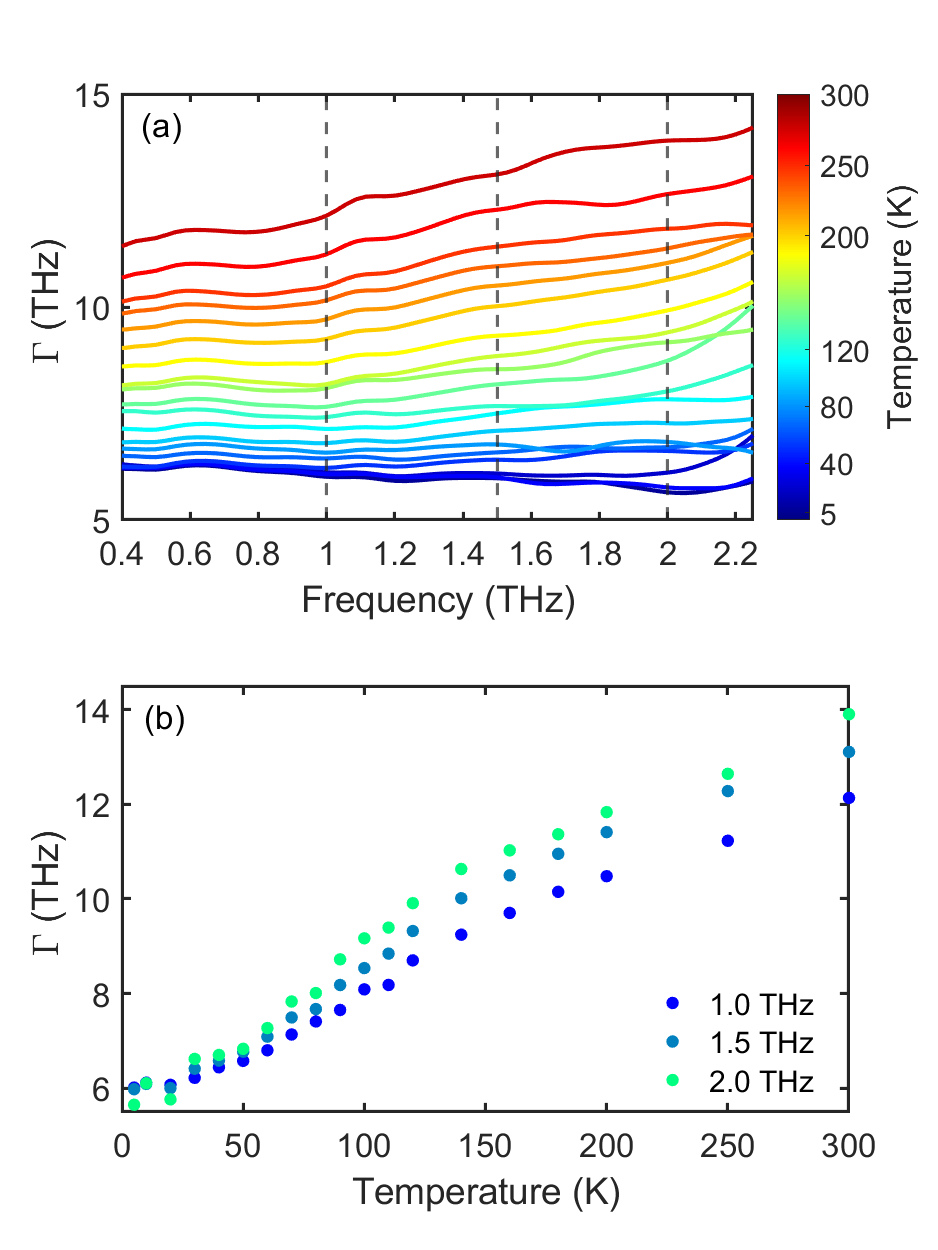}
		\caption{Extended Drude scattering rate ($\Gamma = 2 \pi/ \tau$) of YbAl$_3$ as a function of frequency (a) and temperature (b) taken along frequency cuts indicated by vertical dashed lines in the upper figure.}
		\label{EDM_scatrate}
\end{figure}

\section{Discussion}

\begin{figure}[h]
		\includegraphics[width= \columnwidth]{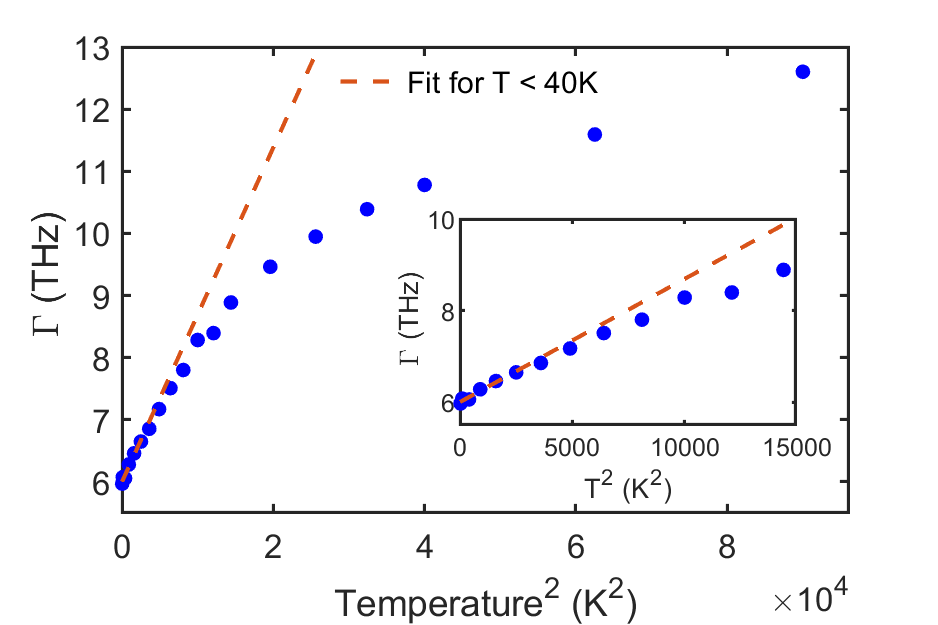}
		\caption{Extended Drude scattering rate of the YbAl$_3$ sample at 1 THz plotted against $T^2$ to compare to FL predictions. Note the minor deviations of the experimental data from the $\Gamma \propto T^2$ linear fit for $T < 40$ K up to 100 K, suggestive of a slow crossover above $T^*$. }
		\label{EDM_scatrateT^2}
\end{figure}

\begin{figure}[h]

		\includegraphics[width= \columnwidth]{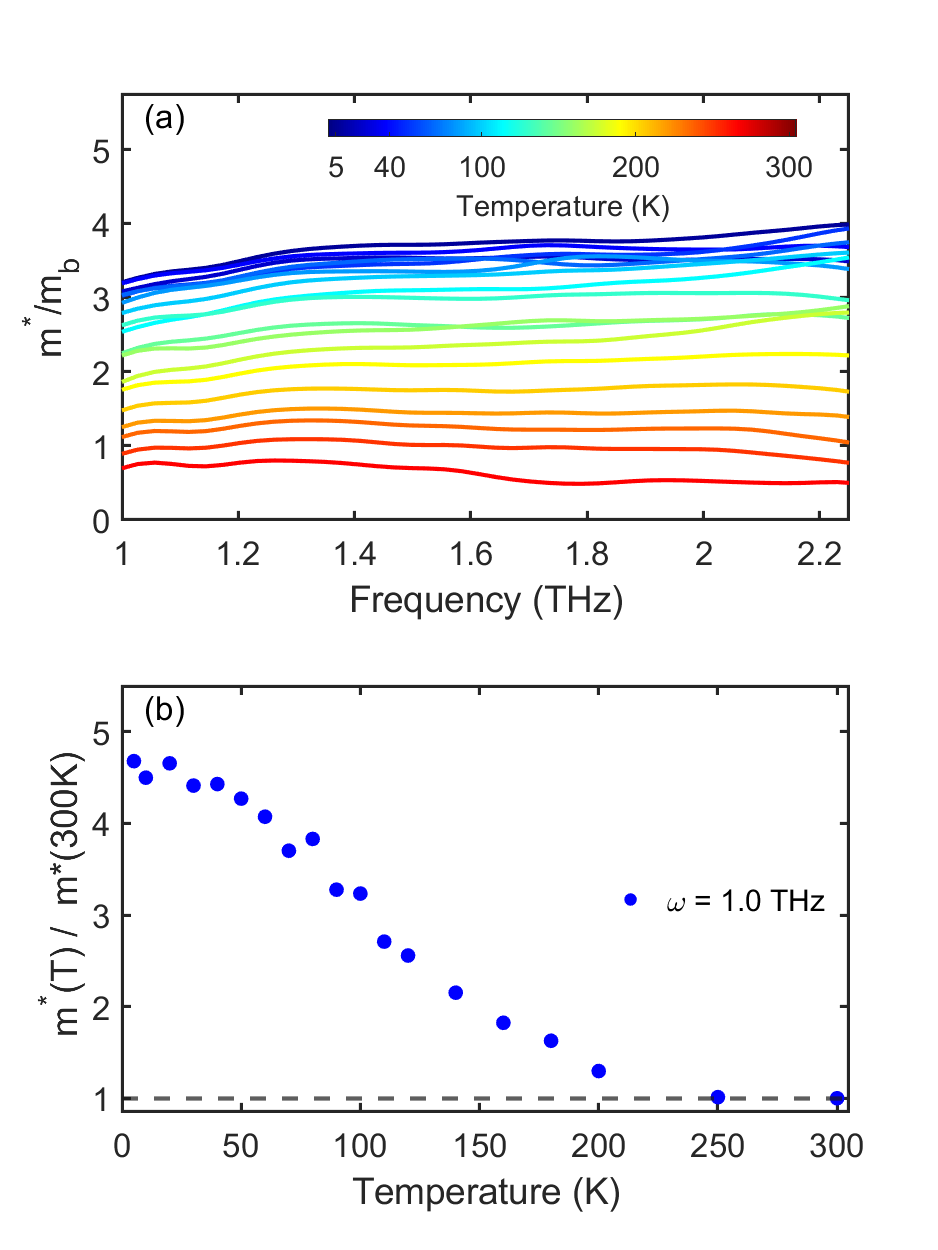}
		\caption{Extended Drude effective mass of the YbAl$_3$ sample as a function of frequency (a) and temperature (b) taken as a frequency cut at $\omega = 1.0$ THz. The effective mass demonstrates a sharp decline below 1 THz which arises from a phase-sensitive artifact. }
		\label{EDM_meff}
\end{figure}

From the comparison of the optical response of LuAl$_3$ to YbAl$_3$, it appears that the full $f-$orbital in the Lu compound prevents the development of the heavy quasiparticles as seen in the YbAl$_3$ film. In order to address the complex quasiparticle behavior, it is useful to use the framework of the extended Drude model by allowing the scattering rate and effective mass to take on both frequency and temperature dependence \cite{Allen1977}. These quantities can be arrived at in a model-free fashion by inverting the complex conductivity as, 

\begin{eqnarray}
\frac{m^*(\omega)}{m_b} = -\frac{\omega_p^2}{4 \pi} \frac{1}{\omega}\; Im \left[ \frac{1}{\widetilde{\sigma}(\omega)}\right] , 
\\
\frac{1}{\tau(\omega)} = \frac{\omega_p^2}{4 \pi} \; Re \left[ \frac{1}{\widetilde{\sigma}(\omega)}\right] , 
\end{eqnarray} 
where $m_b$ is the band mass and $\omega_p$ is the intraband plasma frequency. The exact value of the plasma frequency only serves to scale the renormalized mass and scattering rate and has no impact on their temperature and frequency scaling.

The intraband plasma frequency was determined by adding the spectral weight of each of the fitted low-frequency Drude oscillators. In YbAl$_3$, since the high frequency Drude-Lorentz oscillator was centered at 2800 cm$^{-1}$, our parameterization gave clear separation between the intraband and interband oscillators. Hence, the intraband plasma frequency was determined to be $\omega_p = 2 \pi \times (95 \pm 10)$ THz. A plot of the temperature dependent spectral weight $\omega_p^2$ for both YbAl$_3$ and LuAl$_3$ is shown in Fig.~\ref{MIR_wp2}. The LuAl$_3$ film demonstrated a small decrease in the spectral weight with temperature.  In YbAl$_3$ a small decrease was also observed as the temperature decreased below 50K. These decreases in the spectral weight can be explained as a change in the carrier density $n$ or the effective band mass.

In Fig.~\ref{EDM_scatrate} we present the extended Drude scattering rate as both a function of frequency and temperature. At low temperature frequency dependence of the scattering rate does not exhibit a large frequency dependence within the THz spectral range. Examining the temperature dependence of the scattering rate taken along different frequency cuts, it is clear that the scattering rate becomes enhanced as the temperature is raised above 50 K and begins to show signs of saturation up to 300 K.Due to the phase space constraints on the quasiparticles scattering in the Fermi liquid regime, the scattering rate is expected to demonstrate $\omega^2$ and $T^2$ dependence. Nevertheless, on energy scales exceeding the correlation scale $k_b T^* \approx 3.4$ meV, the quasiparticle description can begin to breakdown. It is also interesting to note that the frequency dependence that develops at higher temperature is behavior beyond Matthiessen's rule as in a conventional Fermi liquid, one would expect T$^2$ and $\omega^2$ dependences to add.

In Fig.~\ref{EDM_meff} the extended Drude effective mass is presented. Below 1 THz, the effective mass seems to dip suddenly; for higher frequencies no significant frequency-dependence is demonstrated. However, when taking a frequency cut at 1 THz we observe a gradual mass enhancement of approximately a factor of 4 is found as the temperature is lowered. The magnitude of the effective mass renormalization is observed to be less than found from other experimental methods \cite{Okamura2004, Ebihara2000}.   For instance, \textcite{Okamura2004} found an approximately factor of 6 change from room temperature to 8K, when considering the lowest frequencies.

It is reasonable to expect that strong disorder scattering interferes with the formation of the mass enhanced state. While we observe the well-defined FL state and the slow crossover phenomena in this YbAl$_3$ film the reason for the smaller mass enhancement seen is unclear. One possibility is that coherent Kondo scattering at low temperatures in these relatively low residual resistivity ratio (RRR) films is strongly affected by the presence of impurities in the lattice \cite{Hamidian2011,Kumar2014,Costa2019}. 

From the extended Drude analysis, it is clear that the behavior of the scattering rate and effective mass indicate that the crossover into the FL regime is not particularly sharp. Unlike other canonical heavy-fermion compounds which can show a dramatic increase in the effective mass~\cite{Stewart1984,Degiorgi1999}, the weaker mass enhancement of quasiparticles and the slow crossover as a function of temperature has been proposed to originate from the mixed-valence states of Yb \cite{Cornelius2002}. The two important energy scales in this material are set by the FL coherence temperature $T^* = 40$ K and the Kondo temperature $T_K = 670$K. \citeauthor{Lawrence2001} found the existence a slow crossover of the in YbAl$_3$ and related YbXCu$_4$ compounds (X = Ag, Cd, In, ...) from transport measurements. Based on results from a more recent photoemission study on the same class of YbXCu$_4$ compounds, the phenomena of slow crossover between the low temperature and the high temperature normal state was found consistent with residual low energy spectral weight that persists into intermediate temperatures \cite{Anzai2020}. This residual spectral weight phenomena is consistent with what we observe in YbAl$_3$ as seen by the enhanced effective mass above the FL coherence temperature $T^*$. This suggests that in YbAl$_3$ hybridization effects are still present above $T > 40$K and demonstrate the slow crossover in YbAl$_3$ as driven by protracted screening processes predicted by the PAM. 

\section{Conclusion}
The field of heavy-fermion physics is relevant to answering questions regarding exotic phenomena that emerge from correlated electronic systems. The recent development of high-quality MBE thin films have allowed the first TDTS experiments to be conducted on YbAl$_3$ and LuAl$_3$. The low energy scales of the heavy quasiparticles makes TDTS an ideal probe to study quasiparticle dynamics and enhance our understanding of these complex systems. This gives the opportunity for well-characterized systems to be studied carefully at the lowest energy scales. 

From our results, we found that the LuAl$_3$ film was well described by conventional Drude transport. The extended Drude formulism was used to study YbAl$_3$ and the quasiparticle scattering rate demonstrates a similar $T^2$ scaling up to the FL coherence temperature $T^*$ consistent with 4-probe measurements. There was a moderate mass enhancement observed at the lowest temperatures, albeit less than the values suggested by dHvA measurements, specific heat measurements or optical spectroscopy on bulk samples. Nevertheless, there is still evidence from the temperature dependence of extended Drude scattering rate $\Gamma \sim T^2$ for the existence of the low-temperature FL state that is consistent with previous studies on bulk crystals.  The reason for the smaller mass enhancement is unclear, but it could be related to disorder scattering in these relatively low RRR films. Regardless, the mass renormalization extracted from the extended Drude analysis gives evidence for a slow crossover in the mass renormalization as found in other Yb-based mixed valence and heavy fermion compounds. At the lowest temperatures, $\Gamma\sim T^2$ is followed up to 40 K, but the experimental data follows the $T^2$ trend closely up to intermediate temperatures $\sim 100$ K.

\textit{Acknowledgements:} The work at J.H.U was supported by the Moore Foundation EPiQS Grant No.(90088577). The work at Cornell was supported by the Air Force Office of
Scientific Research Grant No. FA9550-15-1-0474 and by the National Science Foundation (Platform for the Accelerated Realization, Analysis, and Discovery of Interface Materials, PARADIM) under Cooperative Agreement No. DMR-1539918. Substrate preparation was performed in part at the Cornell NanoScale Facility, a member of the National Nanotechnology Coordinated Infrastructure (NNCI), which is supported by the NSF (Grant No. ECCS-1542081, DMR-170925).

\bibliography{YbAl3}

\end{document}